\documentclass[journal=nalefd,manuscript=letter]{achemso}

\usepackage{caption}
\usepackage{subcaption}
\usepackage{multicol}
\usepackage{multirow}

\newcommand{\onlinecite}[1]{\cite{#1}\hspace{0 ex}} 
\author{Dimitars Jevtics}
\affiliation{Institute of Photonics, SUPA Department of Physics, University of Strathclyde, Glasgow, United Kingdom}

\author{John McPhillimy}
\affiliation{Institute of Photonics, SUPA Department of Physics, University of Strathclyde, Glasgow, United Kingdom}

\author{Benoit Guilhabert}
\affiliation{Institute of Photonics, SUPA Department of Physics, University of Strathclyde, Glasgow, United Kingdom}

\author{Juan A. Alanis}
\affiliation{Department of Physics and Astronomy and Photon Science Institute, The University of Manchester, Manchester, United Kingdom}

\author{Hark Hoe Tan}
\affiliation{Department of Electronic Materials Engineering, Research School of Physics, The Australian National University, Canberra, ACT, Australia}

\author{Chennupati Jagadish}
\affiliation{Department of Electronic Materials Engineering, Research School of Physics, The Australian National University, Canberra, ACT, Australia}

\author{Martin D. Dawson}
\affiliation{Institute of Photonics, SUPA Department of Physics, University of Strathclyde, Glasgow, United Kingdom}

\author{Antonio Hurtado}
\affiliation{Institute of Photonics, SUPA Department of Physics, University of Strathclyde, Glasgow, United Kingdom}

\author{Patrick Parkinson}
\affiliation{Department of Physics and Astronomy and Photon Science Institute, The University of Manchester, Manchester, United Kingdom}

\author{Michael J. Strain}
\email{michael.strain@strath.ac.uk}
\affiliation{Institute of Photonics, SUPA Department of Physics, University of Strathclyde, Glasgow, United Kingdom}
\title[]
{Characterisation, Selection and Micro-Assembly of Nanowire Laser Systems}


\begin{document}
	
		
		%
		%
		%
		

	

	\begin{abstract}
		Semiconductor nanowire (NW) lasers are a promising technology for the realisation of coherent optical sources with extremely small footprint. To fully realize their potential as building blocks in on-chip photonic systems, scalable methods are required for dealing with large populations of inhomogeneous devices that are typically randomly distributed on host substrates. In this work two complementary, high-throughput techniques are combined: the characterisation of nanowire laser populations using automated optical microscopy, and a high accuracy transfer printing process with automatic device spatial registration and transfer.  In this work a population of NW lasers is characterised, binned by threshold energy density and subsequently printed in arrays onto a secondary substrate.  Statistical analysis of the transferred and control devices show that the transfer process does not incur measurable laser damage and the threshold binning can be maintained.  Analysis is provided on the threshold and mode spectra of the device populations to investigate the potential for using NW lasers for integrated systems fabrication.
		
	\end{abstract}

	
Nanowire (NW) lasers are ultra-compact, energy efficient sources of coherent light \cite{Huang2001} with potential applications ranging from distributed on-chip sensing \cite{Yang2019, Cui2001} to optical signal processing \onlinecite{Takiguchi2017}.  A significant amount of effort has been made to optimise the growth processes and physical structure of NW devices to improve brightness, control emission wavelength and modal structure \onlinecite{Quan2019,Gniat2019}. Furthermore, a range of schemes have been developed to integrate these lasers with the necessary on-chip optical components to provide a toolbox to produce future systems.  Vertically structured NWs have been directly grown onto silicon waveguide platforms \cite{Kim2017} and discrete NW devices have been transferred to host substrates, post-growth, for integration with waveguides \onlinecite{Jevtics2017}, plasmonics \cite{Sidiropoulos2014,BermdezUrea2017} and complementary NW structures \onlinecite{Zhang2017}.  Typically however, such systems have been proof-of-principle demonstrations of limited numbers of devices, requiring pre-selection of suitable devices and skilled manually controlled micro-assembly techniques. Nevertheless, pick-and-place assembly is a promising route towards automated fabrication of future systems based on NWs.  Future scalable integration of NW devices will have two major requirements, (1) that as-grown NWs can be removed from their growth structure with high yield and (2) that large sets of these devices can be easily characterised before integration to ensure performance matching the application.  The transfer of NWs from their growth substrate to a host wafer usually includes a fracturing stage where devices are physically 'snapped' at some point along their length, producing one of the two reflective facets required for lasing \onlinecite{Alanis2019}. This mechanical process produces populations of devices across a sample with variations in length and facet quality that in turn affect lasing threshold and modal spectrum.  Therefore, in order to progress towards systems incorporating NW laser sources, a scalable approach is required to map the individual device performance and allow subsequent integration of particular devices with high accuracy spatial positioning. Previous work has shown that it is possible to characterise large populations of NWs using automated microscopy based systems \onlinecite{Alanis2017}.  In this work we present the combination of this population measurement method with a spatially resolved pick-and-place technique \cite{Guilhabert2016,Jevtics2017,Xu2018} to allow device binning and transfer.  A randomly distributed population of NW lasers are characterised and binned by their lasing threshold.  The sub-sets of devices are then transferred to a second substrate using computer controlled spatial registration to allow rapid location and sub-micrometre accurate alignment of NWs.  Finally, automated characterisation of the NW devices is repeated on both the original sample and the sample with printed laser arrays to assess the effects of the printing process on device performance. 

The NW lasers are comprised of GaAs-AlGaAs core-shell wires grown on a GaAs substrate using a bottom-up approach, as reported in ref\onlinecite{Saxena2016}.  The devices support multiple transverse mode solutions and have a diameter of $\sim$ 450$\,$nm and a length of $\sim$ 4\,$\mu$m.  Figure 1 shows a schematic flow diagram of the main stages of NW laser processing, measurement and transfer from the growth substrate onto the host samples.  Full process details can be found in the Supporting Information. In summary: (i)-(ii) vertically grown NWs are released and coarsely distributed on a quartz substrate with random orientation. (iii) These devices are then optically measured and spatially mapped to allow device binning by a specific parameter, such as threshold or lasing wavelength. (iv)-(v) Devices from the population are selected for transfer to a second substrate.  An automatically spatially aligned transfer printing process is used to transfer these devices into regular arrays on a second receiver substrate. (vi) Finally, the devices are re-measured using the same setup as (iii).
\begin{figure}[H]
	\centering
	\includegraphics[width=0.72\linewidth]{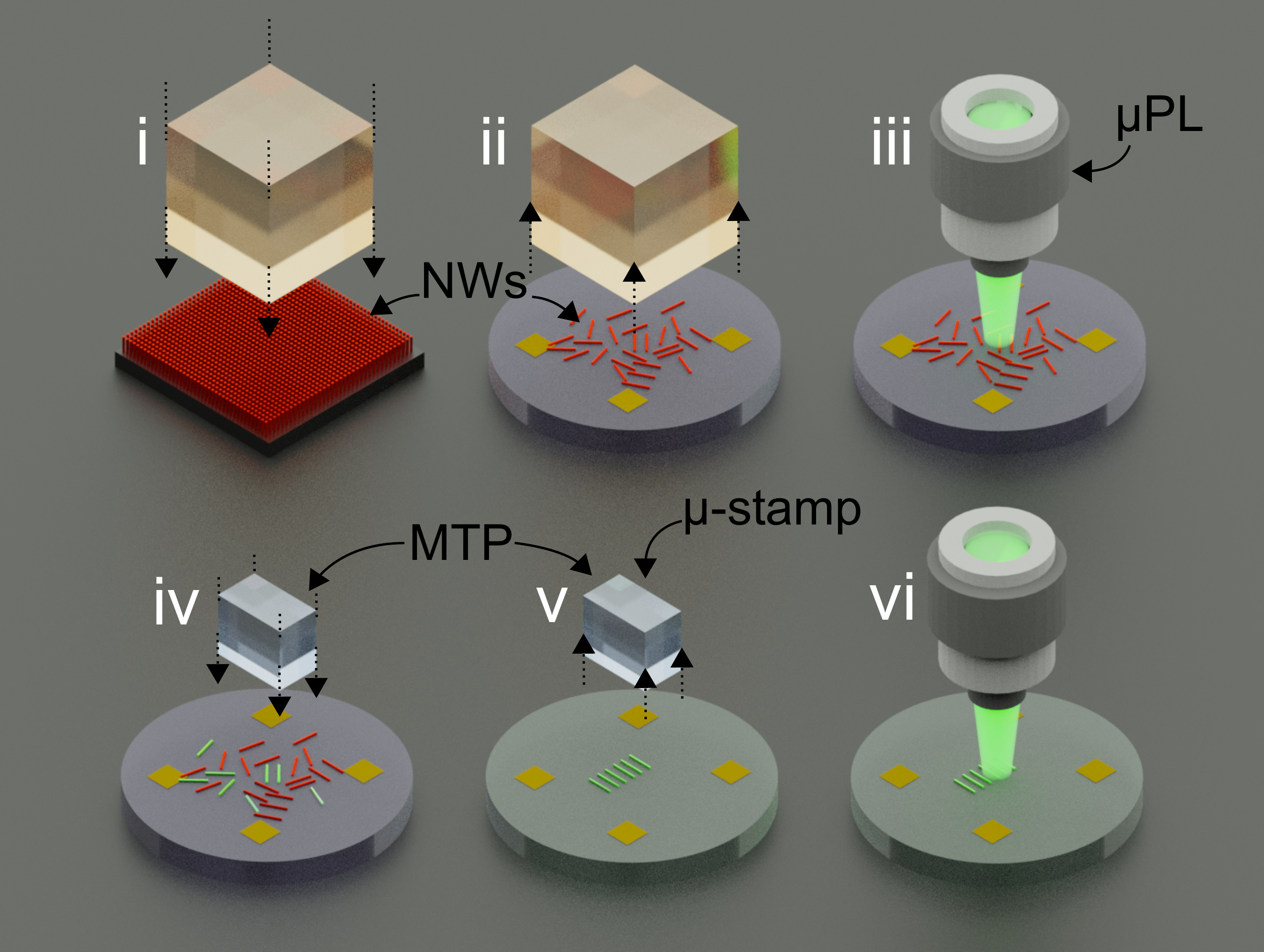}
	\caption{Schematic of the process: (i) NW lasers on their growth substrate are released using a large PDMS block and (ii) transferred onto a target quartz disk and randomly distributed on the surface. (iii) Device populations are characterised using an automated scanning microscopy system and binned by threshold. (iv) Selected NW devices are individually picked using a PDMS $\mu$-stamp and (v) printed onto a second substrate in regular arrays. (vi) Target devices are re-characterised with the system.}
\end{figure}
In the initial step, NW lasers were captured from their growth substrate using a $\sim$ 2 mm$^{2}$ flat polydimethylsiloxane (PDMS) stamp. The large stamp area was brought into direct contact with the vertical array of NWs and translated in one direction in order to induce shearing near the base of the NWs.  The released NWs were bonded to the PDMS surface and can be subsequently transferred to a host substrate.  In this work a z-cut quartz disk, coated with $\sim$ 4 $\mu$m thick photopolymer \cite{NOA65} layer was used as the receiver.  The photopolymer provided an adhesive surface to improve capture of the NWs, and the quartz and photo-polymer are transparent in the visible spectral range, allowing for measurement of the devices in a transmission microscopy arrangement.  The NWs were transferred from the large area PDMS stamp to the quartz sample producing a randomly distributed population of devices on the sample.  The spatial density of these devices was coarsely controlled through the initial release and subsequent printing process parameters.  In this work, a relatively sparse population was targeted to ensure the automatic characterisation method addressed single, easily distinguishable devices.  Figure 2(a) shows an optical microscope image of a random distribution of NWs on the quartz substrate with low spatial density.  Furthermore, both the initial and final quartz receiver samples were patterned with lithographic markers to allow spatial mapping of the NWs to a predefined grid. Previous work on transfer printing of thin membrane devices has shown positional accuracy of the process in the $100$'s of nanometres range \onlinecite{McPhillimy2018}.  
\begin{figure}[H]
	\centering
	\includegraphics[width=0.8\linewidth]{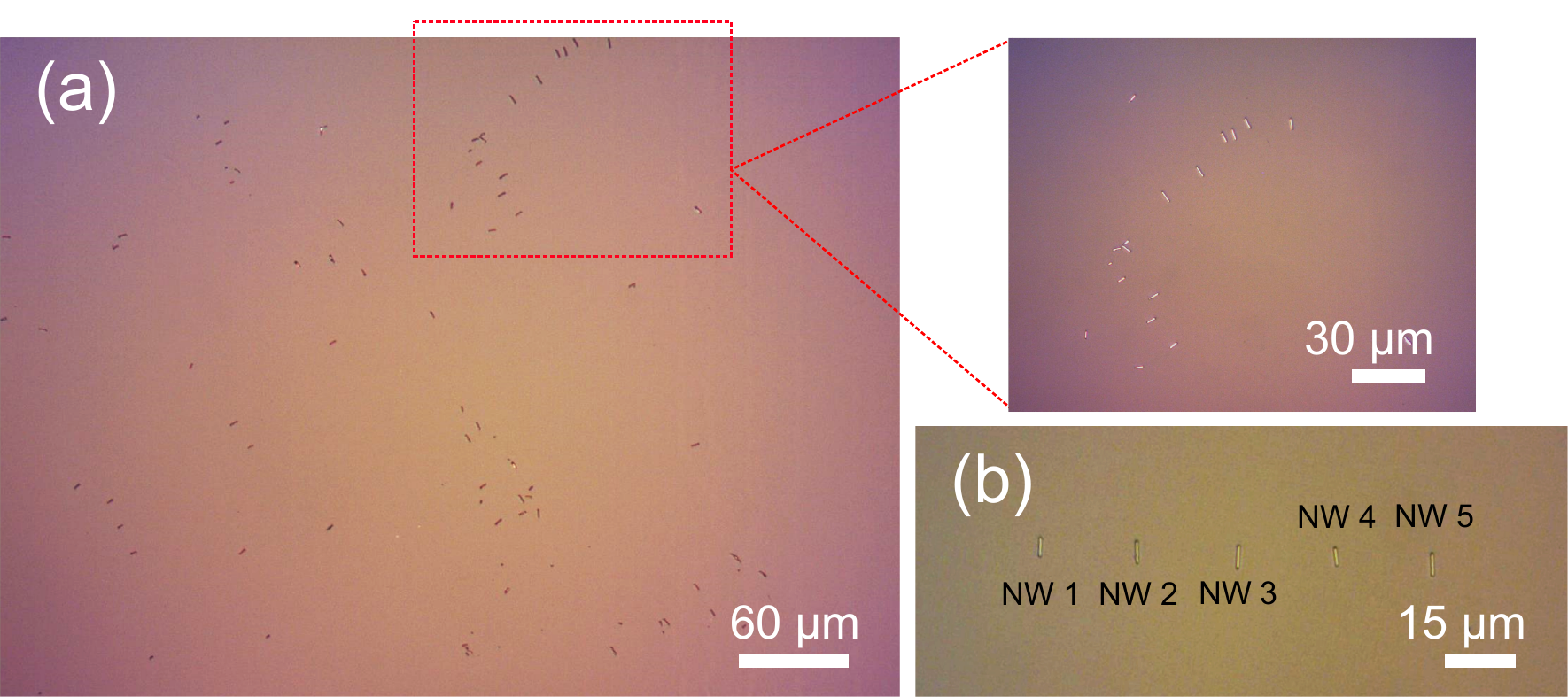}
	\caption{Quartz substrates with printed NW laser devices.  (a) Randomly distributed 1st transfer and (b) 5 device array of binned and deterministically placed devices.}
\end{figure}
The randomly deposited NW lasers were measured using the automated optical characterization setup at room temperature as previously described in ref\onlinecite{Alanis2017}.  Measurement details are presented in full in the Supporting Information. The automated scanning measurement technique allows the direct mapping of NW spatial position to be associated with the optical measurements. The measurements taken include below and above threshold emission spectra, pump energy vs emission power and both dark and bright-field micrographs of the device obtained with a CCD camera \onlinecite{Alanis2017}. In this work we selected the NW laser threshold energy as the binning parameter, as this will be a key metric in the fabrication of future integrated NW systems, designed to operate with similar performance across a chip or to yield lasing emission with comparable pump density across the array.  The initial transfer step created a population of 221 NW devices, out of which 180 of them showed lasing emission, on the first quartz substrate.  Figure 3(a) shows spatial mapping scatter plot of NW lasers across the disk substrate: red dots are coordinates of the NW devices and dark grids depict the alignment markers on the substrate. Figure 3(b) shows a histogram of the initially measured threshold energy densities of the NW laser population.  From this population, five bins of lasing threshold energy were selected as shown in Figure 3(c), varying between 500 - 2500 $\mu$J/cm$^{2}$, with mean values separated by $\sim$ 500 $\mu$J/cm$^{2}$, to produce distinguishable sets.  Five devices were selected per bin, and subsequently printed onto the second quartz sample.  To transfer individual devices, a custom PDMS $\mu$-stamp was used as detailed in ref\onlinecite{Guilhabert2016}.  The selected devices were distributed throughout the population with randomly distributed orientation of their long axes.  The printing process was semi-automated, taking as inputs the position of the NW relative to the registration marker at (0,0) and the orientation taken from the bright field microscopy image of the NW.  The NW target position was defined as a set of coordinates on the target sample, relative to a registration marker on that substrate and all devices were rotated to have their long axes co-linear.  The translation and rotation of the NWs during the printing process was computer controlled based on the measured registration marker and NW device positions. Figure 3(d) shows the measured coordinates of the printed arrays as measured by the characterisation system. Printed devices were separated by $\sim$ 20 $\mu $m to allow individual optical injection and measurement. 

\begin{figure}[H]
	\begin{subfigure}{.45\textwidth}
		\centering
		\includegraphics[width=\linewidth]{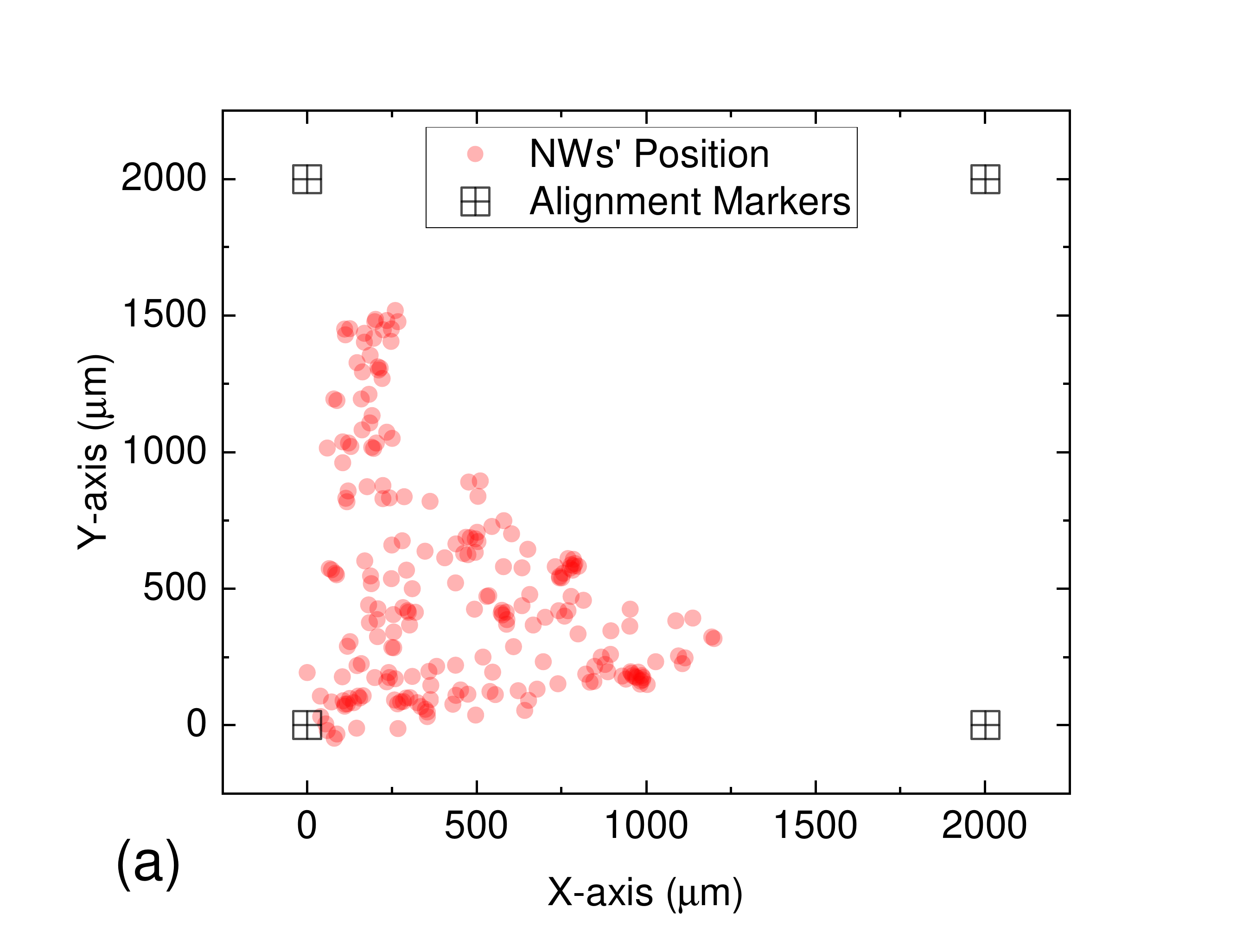}
		\end{subfigure}%
	\begin{subfigure}{.45\textwidth}
		\centering
		\includegraphics[width=\linewidth]{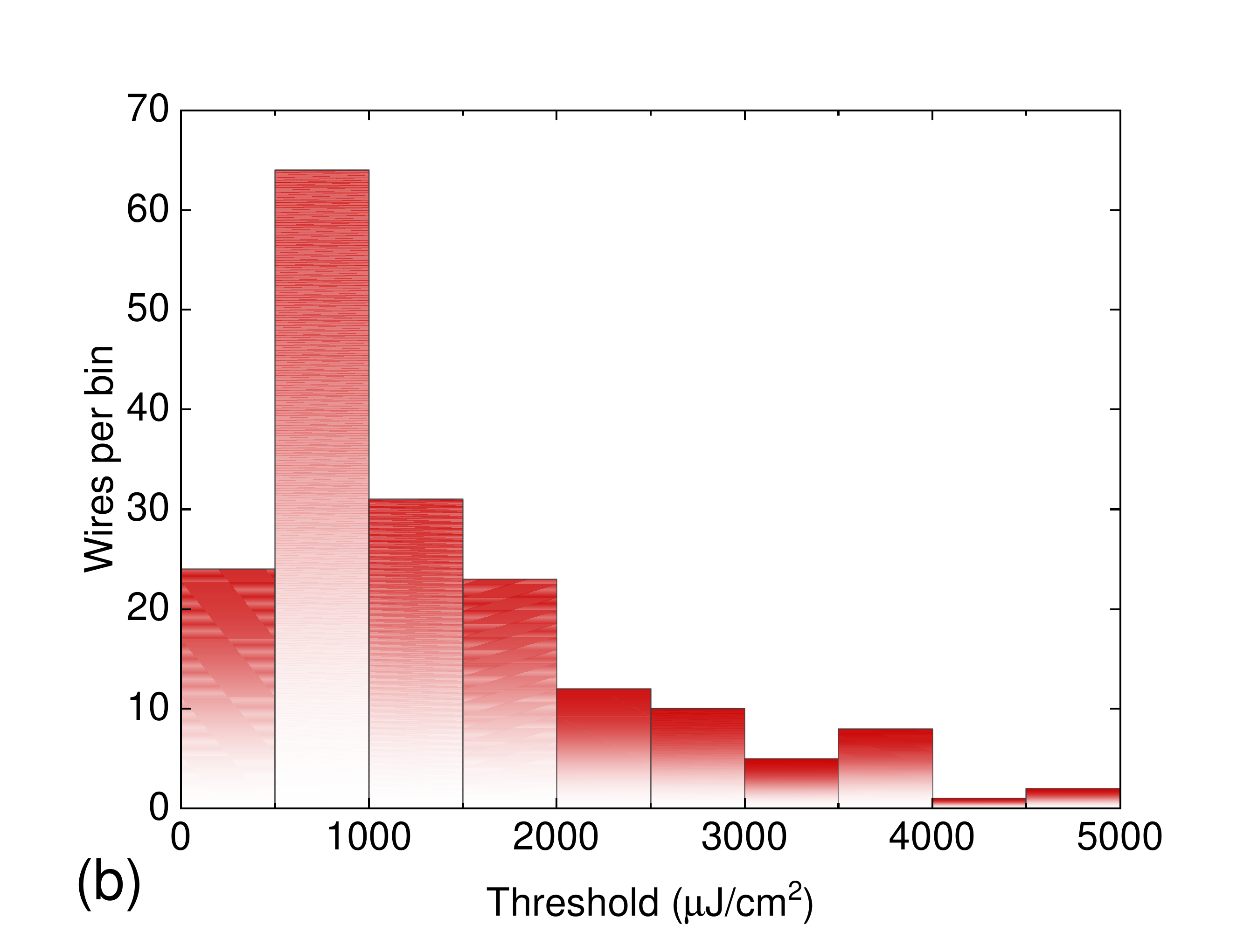}
	\end{subfigure}%

	\begin{subfigure}{.45\textwidth}
	\centering
	\includegraphics[width=\linewidth]{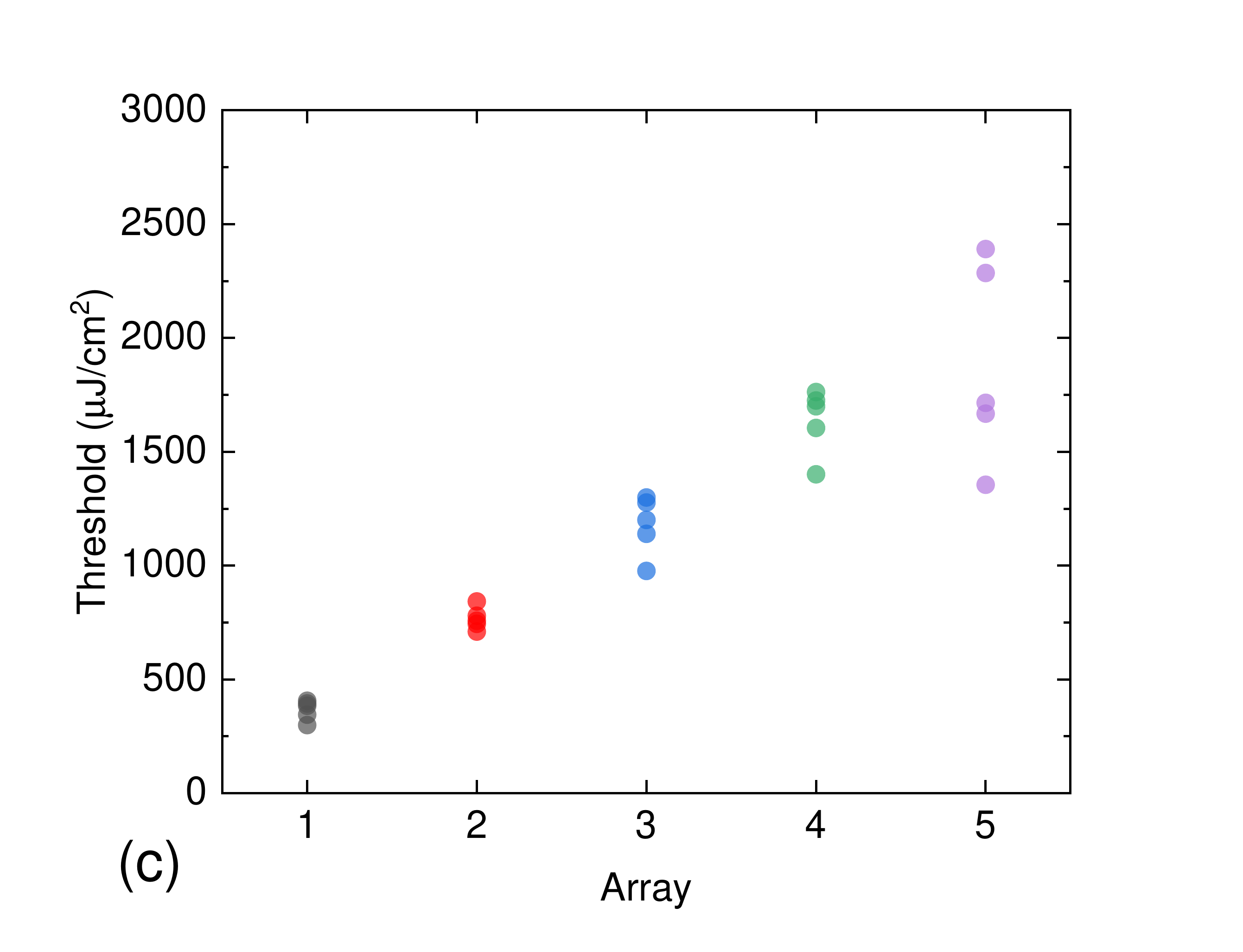}
	\end{subfigure}
	\begin{subfigure}{.45\textwidth}
		\centering
		\includegraphics[width=\linewidth]{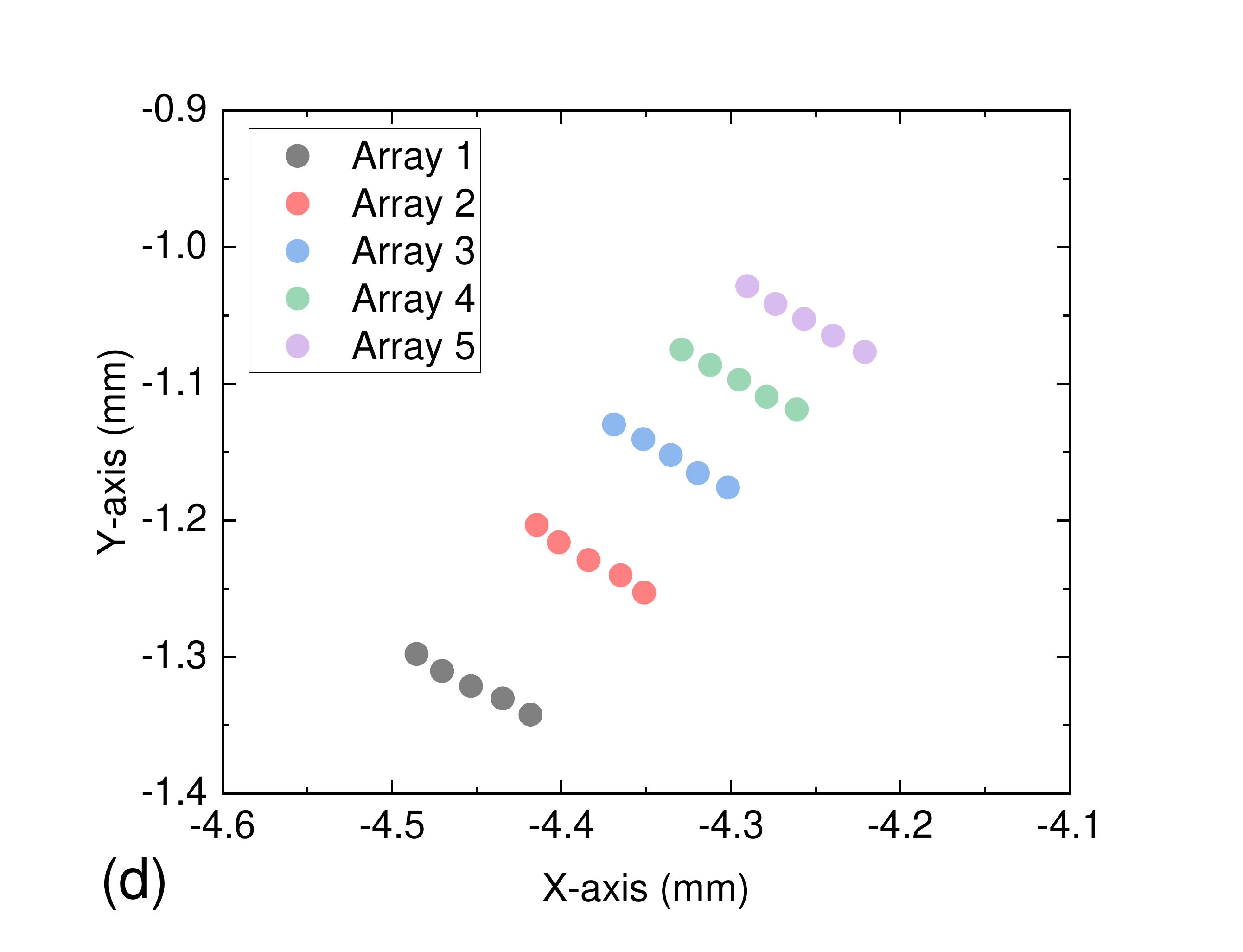}
	\end{subfigure}
	\caption{ (a) Spatial map of 221 NWs (red circles) on quartz substrate identified by the characterisation rig. Dark grids indicate alignment markers on the substrate, with a marker at (0,0) being the reference point. (b) Histogram of measured NW thresholds before the selection and transfer processes.(c) Binning of selected NW lasers by lasing threshold energy. (d) Absolute coordinates of the transfer-printed arrays on the second quartz disk.}
\end{figure}

The post-processing round of measurements was performed on both printed and un-printed samples after a period of six months, during which time both sets of samples were stored together in a desiccator cabinet in a temperature controlled lab. By using the un-printed sample as a reference, we were able to control for variation in experimental conditions (and device aging) between these two measurements. In particular, our comparison allows compensation for changes made to the laser characterisation tool during this period, that may introduce some absolute variation between first and second measurement rounds, but should not affect relative changes between printed and un-printed device sets.
 
 Figure 4 shows the extracted threshold energy densities of the printed (in color) and un-printed (black grids) devices before and after the transfer process, a dashed line is included as a visual aid corresponding to ideal correlation. In the printed device group, 24 out of 25 NW devices retained their lasing emission properties at room temperature, one device did not exhibit lasing. In the case of the un-printed devices, 74 out of the remaining 155 devices exhibited lasing during the second round of measurements.  The reason for the significant decrease in lasing population in the un-printed devices is not clear at this stage, but may be due to thermal effects on the polymer layer at the laser surface during repeated measurements.  
\begin{figure}[H]
	\centering
	\includegraphics[width=0.8\linewidth]{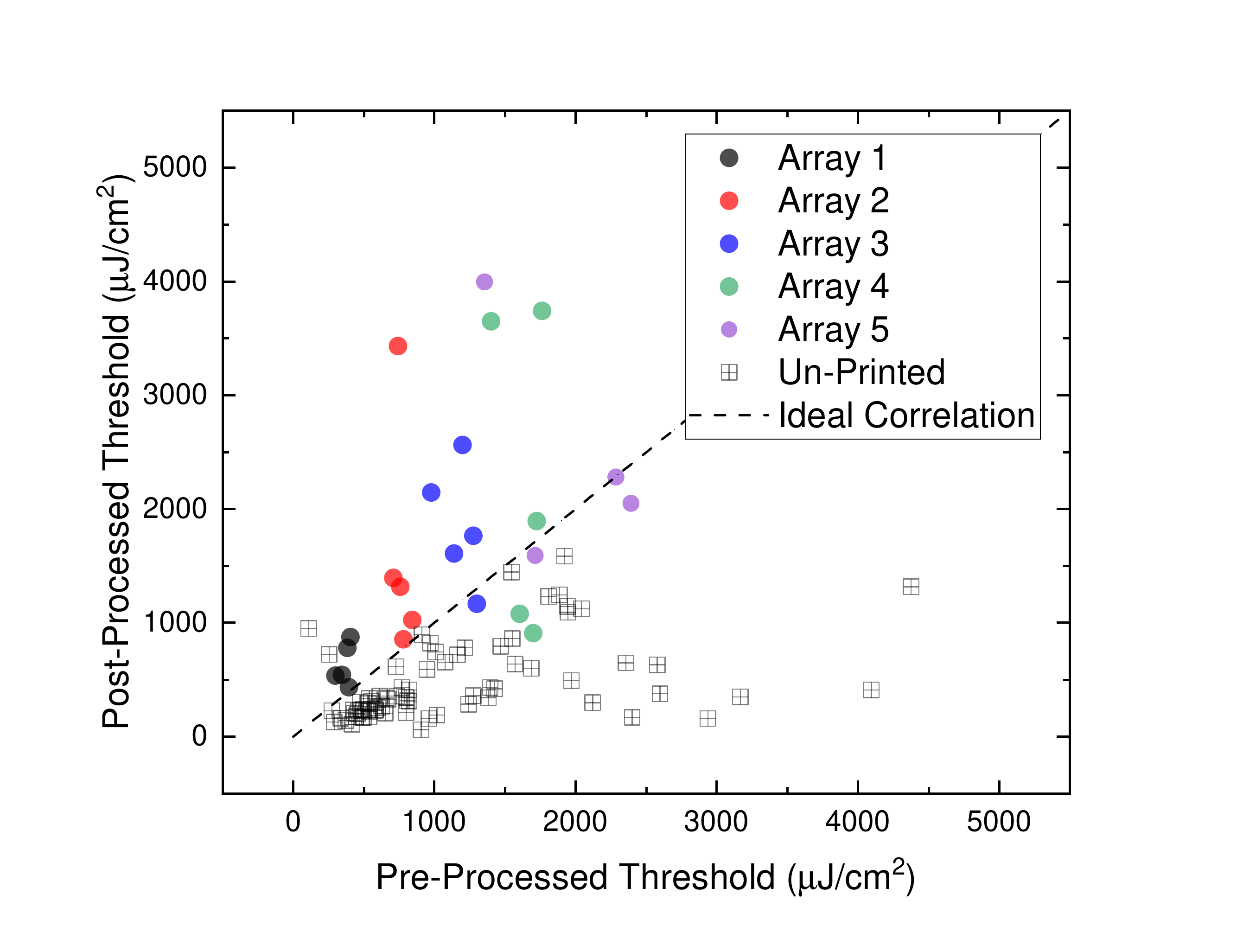}
	\caption{Correlation of laser threshold values between two measurements for printed and un-printed NW lasers. Dashed line represents ideal correlation.}
\end{figure}
Individually, both populations of devices show correlation of their lasing threshold energy density from the first to second measurements, though there are clear differences.  The linear correlation coefficient of threshold for the printed devices is moderate and significant, at 0.462 with \textit{p}-value of .023, showing that the binned devices retain the global trend of threshold ranges, although a clear increase in absolute threshold levels is observed. Similarly, the un-printed devices exhibit a moderate and significant correlation coefficient of 0.435. In this case the second measurement round exhibits a decrease in absolute threshold levels. The threshold energy densities for both populations show systematic variation in absolute values, but retain their global relative trends.  Therefore, the low threshold binned devices can be selected for, even though absolute values are not stable for either printed or un-printed populations, due to changes in the laser excitation system described above.  

As noted earlier, the NW lasers in this work support multiple modes and therefore any changes in the lasing threshold may be due to a change in dominant lasing mode between first and second measurement rounds.  Before considering the lasing spectra, the sub-threshold photo-luminescence of the NWs was measured. As previously shown in ref\onlinecite{Alanis2017}, the photoluminescence spectral form of these devices consists of the two distinct peaks, at $\sim$ 810 and 870 nm, which correspond to the emission from the core and MQW regions of the devices. The shape of emission is unique from wire-to-wire due to the growth inhomogeneity. To verify that the micro-transfer-printing techniques does not affect the printed devices both pre- and post-processing spectra were compared (see Supporting Information). As no significant variations in the PL spectra were apparent, the peak lasing wavelength for both populations of devices was measured, as shown in Figure 5, where red circles represent printed devices and black grids represent un-printed devices.

\begin{figure}[H]
		\centering
		\includegraphics[width=0.8\linewidth]{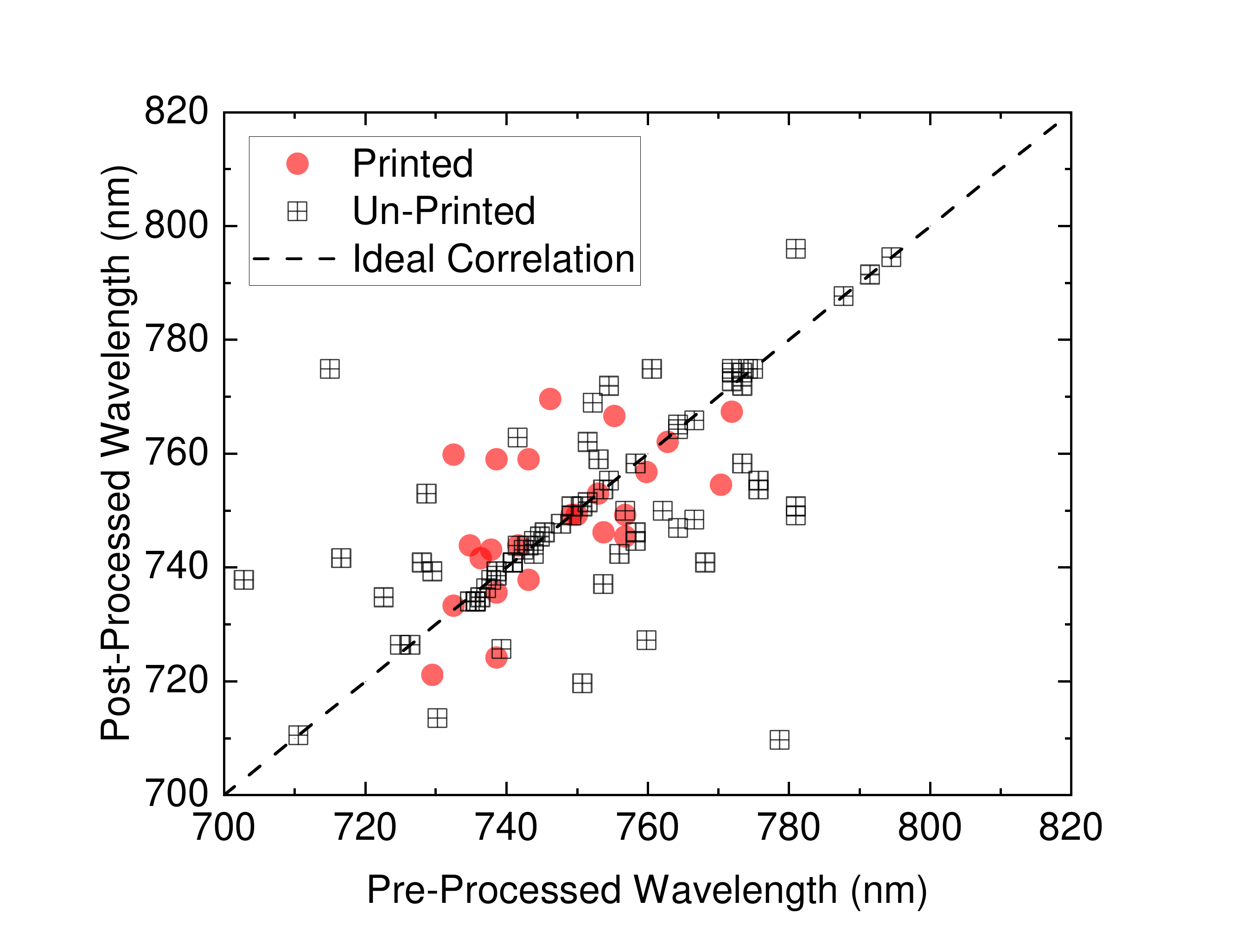}
		\caption{Correlation of peak lasing wavelength between two measurements for printed (red circles) and un-printed (black grids) devices. Dashed line represents ideal correlation.}
\end{figure}

Both populations of devices exhibit similar scatter around the ideal correlation curve, suggesting that the printing process does not dominate variations in peak lasing wavelength. The measured variation in peak lasing wavelength is significant for these multi-mode devices, extending beyond 10 nm in some cases.  The peak wavelength measurement does not give a full account of the laser modal structure and so a further qualitative measure of the spectra can be defined that may help to ascertain if the NW has been physically altered between measurements.  This modal structure of the device is directly related to its physical geometry and so provides a useful probe of any damage induced in the NWs.  Three possible cases are presented in Figure 6 that show how the modal structure of the laser may change between measurements.


\begin{figure}[H]
	\centering
	\includegraphics[width=1\linewidth]{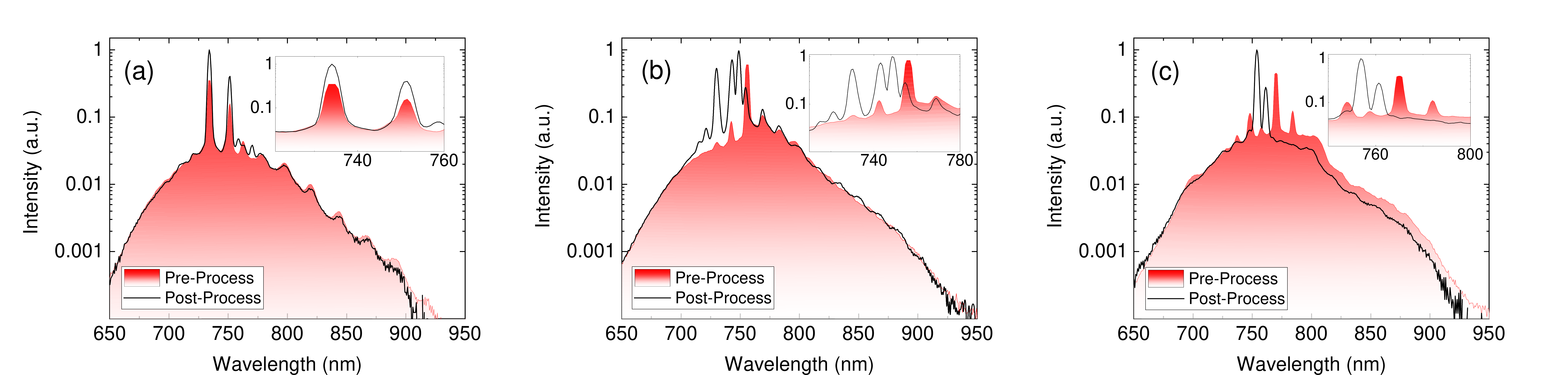}
	\caption{Three representative cases of the printed and un-printed devices which depict state of the spectrum pre- and post-processing: (a) reconstructed, (b) partially reconstructed, (c) non-reconstructed spectrum.}
\end{figure}

In the first case `A', shown in Figure 6(a), the Fabry-Perot (FP) modes of the device in both measurements fully overlap, i.e. the modes overlap in wavelength and exhibit similar distribution of optical power. For the particular device presented, the lasing threshold changed from 734 to 356 $\mu$J/cm$^{2}$, but the modal structure shows very little variation. This suggests that the device losses, gain or absorption properties may be affected but its geometry and hence FP mode structure is consistent. The second common case `B', Figure 6(b), exhibits a spectrum where the FP modes of the device overlap between measurements, but the distribution of optical power has changed significantly. Again, in this case it would suggest that the geometry of the device is unaffected but the internal laser properties, or pumping conditions have changed enough to favor lasing of an alternative mode. Finally, in case `C', in Figure 6(c) the two modal spectra do not appear to overlap well, suggesting that the device has been physically affected, inducing a change in the FP mode structure.  Figures 7 shows the correlation between measured laser threshold values between first and second measurement rounds. The form of each data point marker (black, red and blue) indicates whether the lasing spectral overlap between measurement `A' and `B' resembles case `A', `B' or `C' most closely.

\begin{figure}[H]
	\centering
	\includegraphics[width=1\linewidth]{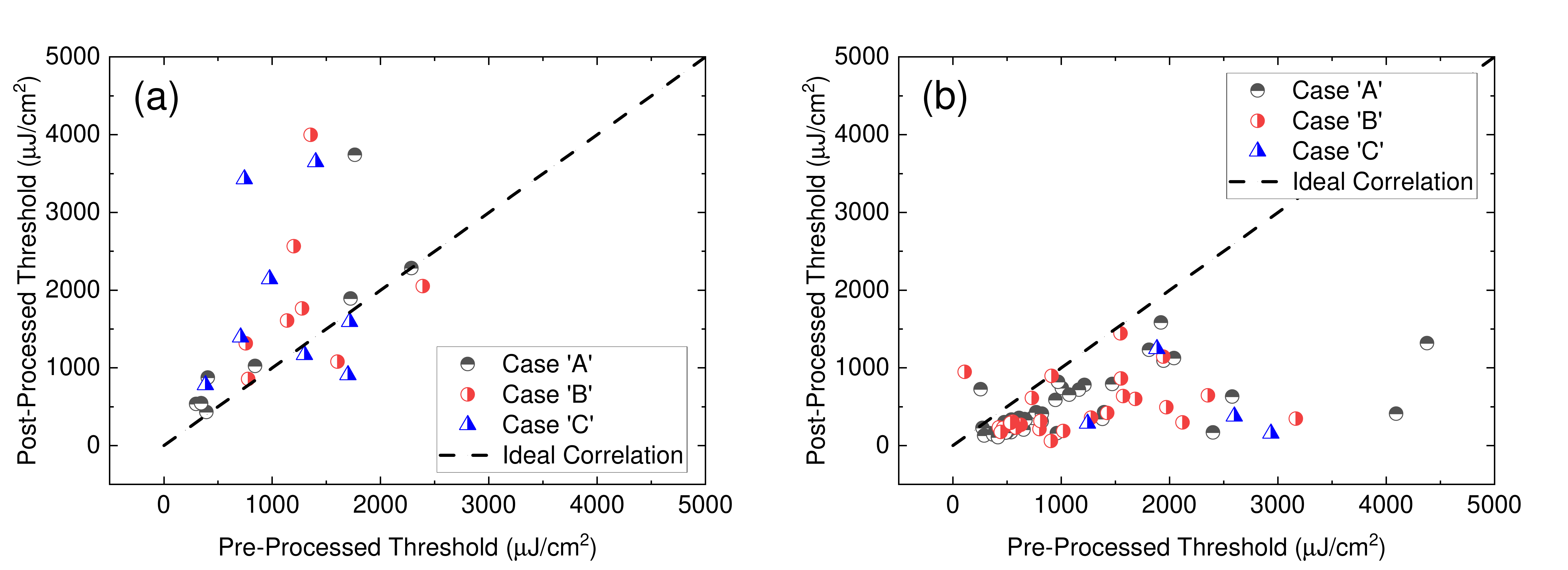}
	\caption{A scatter plot showing modal cases relatively to the threshold change in the NW devices: (a) printed and (b) un-printed devices.}
\end{figure}

Table 1 presents the correlation coefficients and p-values for each of three spectral overlap cases and the full printed and un-printed device populations. In both the printed and un-printed devices, case `A' (full mode overlap) shows good correlation between first and second measurements. Cases `B' and `C' exhibit no significant correlation values for both populations. This suggests that where the device lasing conditions are consistent between first and second measurements, the printing process does not introduce significant detrimental effects to the devices and lasing thresholds are consistent. Nevertheless, the multi-modal nature of the NW lasers studied here allows for variation in lasing mode between measurements and therefore a shift in lasing threshold. In order to maintain tight control over device characteristics single transverse-mode NW lasers could be used.
\\
\begin{table}
	\caption{Threshold current density correlations for devices grouped by modal spectrum correlations}
	\begin{tabular}{lcccc}
		\multirow{2}{5cm}{Mode Spectrum Case [un-printed $\mid$ printed]} & \multicolumn{2}{c}{Un-printed} & \multicolumn{2}{c}{Printed} \\
		\cline{2-5}
		& $\rho$-value & \textit{p}-value  & $\rho$-value & \textit{p}-value\\\hline
		\hline
		All devices [74 $\mid$ 24]   & 0.435 & $<.001$  & 0.462 & .023\\
		Case A [45 $\mid$ 8]  & 0.584   & $<.001$ & 0.841 & .009\\
		Case B [25 $\mid$ 8] & 0.265 & .200 & 0.239 & .568\\
		Case C  [4 $\mid$ 8]& -0.294  & .706 & 0.006 & .989\\
		\hline
	\end{tabular}
\end{table}

In conclusion, we have demonstrated that large populations of NW lasers with random spatial distribution can be rapidly characterised and sorted using their measured lasing parameters. These populations can then be transferred to separate substrates using a semi-automated pick and place technique that uses the same spatial registration markers as the characterisation measurements. Devices can be binned by lasing threshold and spatially arranged into regular arrays on a secondary substrate. Repeated measurements on both un-printed and printed device populations show that the multi-mode nature of the devices allows for variations in measured laser characteristics in both cases. Nevertheless, in cases where the laser modal spectrum is similar in both measurements, the lasing threshold energy densities are strongly correlated, showing that effective binning and device assembly can be achieved. These results suggest that the printing process does not induce any particular detrimental effects in the transferred devices and could be a route to scalable systems assembly from large inhomogeneous distributions of NW devices. Furthermore, by targeting single-mode devices, some of this variability may be mitigated.

\subsection{Methods}
\paragraph{NW laser characterisation:} The as-transferred nanowires were located, imaged and characterised using an automated microscopy platform as previously described \onlinecite{Alanis2017, Alanis2018}. In brief, the sample was moved beneath a 60x objective lens and imaged using a CCD camera. Nanowires were identified on the substrate using machine vision, and each wire was sequentially imaged. A HeNe laser (632.8 nm) at low power was used to excite each wire, and the low-power photoluminescence was collected using a fibre-coupled spectrometer. For the lasing study, an ultrafast laser pulse (250 kHz, $\sim$ 150 fs, 620 nm) was produced using an optical parametric amplifier, with the pulse energy controlled using a computer controlled neutral density filter. The pulse was defocussed to around 800 $\mu $m$^2$ (a circle of 16 $\mu $m$^2$ radius) to uniformly excite the wires. The laser power was continually monitored using a calibrated silicon photodiode. The threshold for each wire was measured by increasing the excitation fluence until a narrow lasing line emerged which was accompanied by an increase in gradient in the integrated emission as a function of excitation energy as described previously \onlinecite{Alanis2018}. The position (relative to the markers), the lasing threshold, the microscopy, the low-power photoluminescence and the lasing spectrum was recorded and stored for every nanowire. The process took approximately 30-60 seconds per nanowire, including threshold measurement.
After transfer, the same process was performed. The un-printed wires were re-located and matched to their initial data using the positional relocation approach outlined in \onlinecite{Parkinson2018}.

\paragraph{Micro-Transfer-Printing:}
Two main forms of transfer printing were used in this work.  In the first case a large PDMS block was used to release NWs from their growth substrate using a shearing motion when the surface of the block was placed in contact with the NW substrate.  The released NWs adhered to the surface of the PDMS block which was subsequently brought into contact with a photo-polymer coated quartz disk.  The block was sheared against the quartz substrate, thereby releasing the NW's.  By controlling the speed of the PDMS translation relative to the quartz disk and applied contact force, the density of released NWs could be roughly controlled.
In the second printing process devices selected during the laser measurement and binning stages were identified with positions defined relative to the on-chip registration marks.  The orientation of the NWs was measured using a bright field microscopy image from the characterisation measurements.  The NWs were then defined relative to the markers in the xy-plane and with a relative rotation of their major axis.  A second set of markers on the receiver substrate allowed positioning of the NWs in a pre-defined grid and with their major axes aligned.  The printing process was carried out using a PDMS microblock with a surface area of $ 10 \times 30$ \,$\mu$m.  This block was aligned with the NW of interest and brought into contact.  The NW was then released from the substrate using a rapid vertical translation of the PDMS block.  The NW was then aligned to the receiver substrate and brought into contact.  A slow release then allowed transfer of the NW to the receiver substrate.  
%
%

\subsection{Author Contributions}
D.J., J.M., B.G., J.A.A. and P.P. carried out the experiments. D.J. and M.J.S. wrote the manuscript with support from all authors. M.J.S., A.H., B.G. and M.D.D. supervised the project. H.H.T. and C.J. fabricated the GaAs-AlGaAs nanowire lasers. A.H., P.P., and M.J.S. conceived the original idea.

\section{Acknowledgments}
We acknowledge the support of EPSRC (EP/P013597/1, EP/R03480X/1) and the Australian Research Council. Access to the nanowire growth facility was made possible through the Australian National Fabrication Facility, ACT Node.

\bibliographystyle{journalstyle}
\bibliography{JevticsNWBinning.bib}

\end{document}